 \documentclass[final,5p,times,twocolumn,nopreprintline]{elsarticle}

\usepackage{amssymb}

\usepackage{hyperref}
\usepackage{booktabs}
\usepackage{graphicx}
\usepackage[table,xcdraw]{xcolor}
\usepackage{multicol}
\usepackage{mdframed}

\journal{Sustainable Computing: Informatics and Systems}

\begin{document}

\begin{frontmatter}



\title{A Comparative Analysis of Energy Consumption  
Between the Widespread Unity  
and  Unreal Video Game Engines}

\author[inst1]{Carlos P\'erez\corref{cor1}}
\cortext[cor1]{Corresponding author. Email address: cperez@usj.es}

\author[inst1]{Javier Ver\'on}
\author[inst1]{Francisca Pérez}
\author[inst2]{M\textordfeminine \'Angeles Moraga}
\author[inst2]{Coral Calero}
\author[inst1]{Carlos Cetina}

\affiliation[inst1]{organization={SVIT Research Group},
            addressline={Universidad San Jorge, Autovía Mudéjar, km. 299},
            city={Villanueva de Gállego},
            postcode={50830}, 
            state={Zaragoza},
            country={Spain}}

\affiliation[inst2]{organization={Alarcos Research Group},
            addressline={University of Castilla-La Mancha, Paseo de la Universidad, 4}, 
            city={Ciudad Real},
            postcode={13071}, 
            state={Ciudad Real},
            country={Spain}}

\begin{abstract}
The total energy cost of computing activities is steadily increasing and projections indicate that it will be one of the dominant global energy consumers in the coming decades. 
However, perhaps due to its relative youth, the video game sector has not yet developed the same level of environmental awareness as other computing technologies despite the estimated three billion regular video game players in the world.
This work evaluates the energy consumption of 
the most widely used industry-scale video game engines: Unity and Unreal Engine. Specifically, our work uses three scenarios representing relevant aspects of video games (Physics, Static Meshes, and Dynamic Meshes) to compare the energy consumption of the engines. The aim is to determine the influence of using each of the two engines on energy consumption.
Our research has confirmed significant differences in the energy consumption of video game engines: 351\% in Physics in favor of Unity, 17\% in Static Meshes in favor of Unity, and 26\% in Dynamic Meshes in favor of Unreal Engine. These results represent an opportunity for worldwide potential savings
of at least 51 TWh per year, which is equivalent to the annual consumption
of nearly 13 million European households. This might encourage a new branch of research on energy-efficient video game engines.

\end{abstract}




\end{frontmatter}


\section{Introduction}
\label{sec:Introduction}

An increasing number of human activities, even those traditionally associated with the physical world, rely to a large extent on software for their development. A paradigmatic example is the automotive industry, where a significant part of the value chain is moving to software, a trend that the current electrification of the industry is likely to accelerate. The transformation of traditional industries through software is happening on a global scale.

This massive deployment of software in all of our activities is accompanied by a sharp increase in the energy costs required to run it. The total energy cost of computing activities is steadily increasing, and projections indicate that it will be one of the dominant global energy consumers in the coming decades~\cite{AndraeEdler}. Therefore, there is growing interest in understanding the environmental impact of software in the various activities in which it is relevant and how this impact can be evaluated and decreased.

One of the sectors where the growth has been most evident in recent years is video games. It is estimated that there are more than three billion regular video game players in the world, whether on PCs, mobile devices, dedicated consoles, or other types of devices~\cite{Statista2023, ExplodingTopics2023}. The development of many recent video games is characterized by, among other things, greater graphical realism and an increase in the number of interactions and players, which requires more powerful equipment capable of managing this complexity without degrading the user experience. All of these factors suggest that the energy consumption of video games and their environmental impact are becoming increasingly important.

We have estimated the global energy consumption of video games per year to range between 230 TWh (just considering PC gamers) and 347 TWh (including gaming consoles). The estimation was made by measuring the average power required when playing \textit{Baldur’s Gate III}, one of the most popular PC games released in 2023~\cite{clement2023most}. This was 358.6~W for just the video game execution (removing the consumption by the PC background processes). The estimation assumes all gamers played the same game.
A recent global estimation of average weekly hours spent playing was 8.45 hours~\cite{clement2023average}, and the global estimation of people who played video games was 3.38 
billion;~\cite{newzoo2023global} 43\% of the total players are estimated to play on PC and 65\% are estimated to play on PC or console~\cite{newzoo2023consumers}. Gaming consoles are supposed to be similar to gaming PCs in terms of power, and mobile devices are not considered in the estimation, as their power is typically one or two orders of magnitude smaller. 


To put these numbers into perspective, the average annual amount of electricity purchased by a EU member residential electric-utility customer was 3900~kWh\cite{EUdwellings}.
 Therefore, those figures are equivalent to the annual energy consumption of approximately between 59 million (just PC gamers) and 89 million households (including gaming consoles) in the European Union. This does not take into into account the consumption of the additional infrastructure beyond the player's computer that games like \textit{Baldur's Gate III} require; e.g., servers for multiplayer gameplay. This also does not count the new trend of cloud gaming (e.g., GeForce Now), where the game is run on remote servers and then played on the local device, which can increase the overall energy consumption~\cite{newzoo2023global}. 

However, perhaps due to its relative youth the video game sector has not yet developed the same level of environmental awareness as other computing technology sectors. In the case of computing technologies in general, there are research efforts in the area of green cloud computing~\cite{atrey13}, green mobile computing~\cite{6085386}, green software~\cite{Pereira2021}, and green data centers~\cite{wang2011} among others. Nevertheless, in the case of video games, efforts are specifically concentrated on what is called Green-by computing.
There are two perspectives for analyzing the sustainability of software engineering in general and its energy consumption in particular: Green-in and Green-by~\cite{gutierrez2023green}. The first one deals with improving the energy consumption of a field per se, e.g., video games, whereas the second one is focused
on the application of a field (video games) to improve sustainability in any context. For example, \cite{johnson2017gamification} conducted a systematic review to assess the effectiveness
of gamification and serious games in impacting domestic energy consumption.

Modern video games are highly complex products of software that rely on interrelated technologies to function. It is common to use a video game engine to build video games. The video game engine enables developers to create the different games in a much more agile way, without having to repeat or code low-level elements whose design from scratch would be very costly and incompatible with the pace of development of these products. The most well-known and widespread video game engines are Unity and Unreal Engine~\cite{unitygaming2022}~\cite{activeplayer2023fortnite}. Comparable video games in terms of  visuals and physics can be achieved with both Unity and Unreal
engine; however, the developers do not know the impact of the
engines in terms of energy consumption because it has not been
explored so far.



This work constitutes a first approach to the evaluation of energy consumption 
of a video game using the two most widely used industry-scale video game engines: Unity and Unreal Engine. Specifically, our work uses three scenarios representing relevant aspects of video game development (Physics, Static Meshes, and Dynamic Meshes) to compare the energy consumption of the engines. The aim is to determine the influence of using each of the two engines on energy consumption. The results show that there are significant differences between the versions coded with Unity and Unreal Engine, depending on the scenario. These results
can be revulsive for the industry, triggering a race in which
manufacturers will make their engines better in terms of efficiency,
approaching the best values identified.
Furthermore, these results might encourage a new branch of research on energy-efficient video game engines.

The paper is structured as follows: Section~\ref{sec:Overview VG} provides a concise summary of video game engines: what they are and how they are used within a video game. Section~\ref{sec:MeasEner} discusses previous work on assessing software energy consumption, and Section~\ref{sec:ExpDescr} explains the experiment in detail. After introducing the fundamental aspects of the experiment, section~\ref{sec:ExpProc} delves into the examination of the procedure required for its execution.
The threats to the validity of the experiment are explained in Section~\ref{sec:threatsToValidity} and, lastly, conclusions are summarized and future work is outlined in Section~\ref{sec:Conclusion}.

\section{A brief overview on video game engines}
\label{sec:Overview VG}

Video game engines, often simply referred to as "game engines", are software products designed to ease the development of video games. A game engine provides game developers with essential tools and functionalities, enabling them to focus on the creative and unique aspects of the game rather than having to program all of the basic functionalities from scratch. The functionalities that a game engine usually implements are the following:

\begin{itemize}
    \item Rendering engine: This engine is responsible for the graphical display and renders game elements on the screen. It employs advanced computer graphics techniques, including 2D and 3D rendering, lighting, and shading.
    \item Physics engine: This engine simulates the laws of Physics within the game world, adding realism to the movement and interactions of objects. This includes the simulation of gravity, collisions, and other forces.
    \item Sound system: This component manages all aspects of audio in the game, which is essential for including music, sound effects, and voice acting. It aids in creating an immersive experience for the player and in providing feedback for the player’s actions.
    \item Input system: This component handles how the game is going to receive the user input from the different external devices supported by the system (e.g., mouse, keyboard, or controllers).
    \item Game logic: This component handles how the game is going to work and be played, including scripting systems.
\end{itemize}

 Game engines usually include level editors, graphical user interface (GUI) editors, animation systems, and profiling tools; allowing developers to design, test, and polish  their games faster. A game engine also usually offers other functionalities that greatly shorten the development time, such as cross-platform support, which enables the game to run on multiple devices and platforms just by building it for different target devices.

Most game developers use game engines, (60\% as of 2022~\cite{slashdata2022}). There are multiple industry-scale engines, but Unity has the largest share of the game engine market: 38\% of game developers who use game engines use Unity as their primary engine~\cite{slashdata2022}. The next most popular game engine, Unreal Engine, has 15\% usage as a primary engine~\cite{slashdata2022}. Normally, Unreal Engine is the preferred choice for high budget developments. In fact, nearly 80
of the most highly anticipated games released in 2023 are powered by Unreal Engine~\cite{epic2023roundup}.

On the one hand, Unity is very flexible in performance depending on the target platform, allowing the creation of mobile games and XR games to usually be easier than Unreal Engine. It uses C\# as scripting language and is considered to be easier to use by beginners because of its user-friendly interface, and because it usually can get as complex as needed. 

On the other hand Unreal Engine is renowned for its advanced graphics and high-fidelity rendering capabilities. It offers C++ as scripting language, and it also offers Blueprint Visual Scripting. This visual scripting system allows developers to create game logic and behaviors using a node-based interface without traditional programming to prototype or create game logic and interactions without deep programming knowledge. However, even with the visual scripting, Unreal Engine is considered to have a steeper learning curve than other alternatives such as Unity. 

\begin{figure*}[h]
\centering
\includegraphics[width=.75\textwidth]{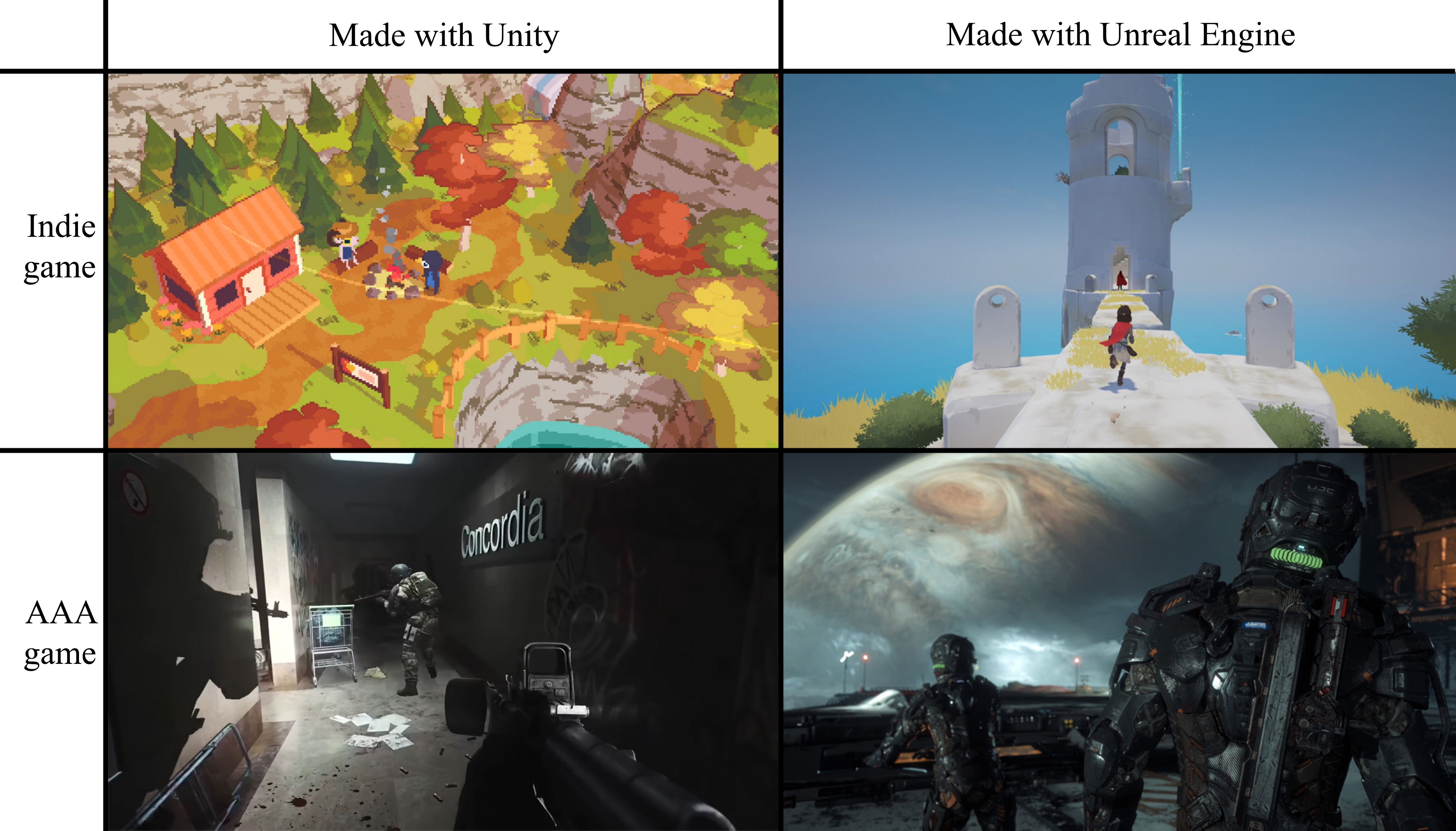}
\caption{Indie and AAA games made in Unity and Unreal Engine: \textit{A Short Hike} (indie, Unity), \textit{RiME} (indie, Unreal Engine), \textit{Escape from Tarkov} (AAA, Unity), and \textit{The Calisto Protocol} (AAA, Unreal Engine).}
\label{fig:indieAndAAAGames}
\end{figure*}

 
On an industry scale, both Unity and Unreal Engine have been used to develop indie and AAA games. 
"Indie" is short for "independent". An indie game is typically developed by solo developers or small teams without the financial support for a large development. Indie games are known for their innovative and creative design.
For example, \textit{A Short Hike} is an indie game developed with Unity and \textit{RiME} is an indie game developed with Unreal Engine. Both games are shown in Figure~\ref{fig:indieAndAAAGames}.
AAA (also called "triple-A") games are the video game equivalent of blockbuster movies. They are high-budget, high-profile games that are typically produced and distributed by large, well-established gaming companies. AAA games are known for their impressive visuals and high production values.
For example,  \textit{Escape from Tarkov} is a AAA game developed with Unity and \textit{The Calisto Protocol} is a AAA game developed with Unreal Engine. Both games are also shown in Figure~\ref{fig:indieAndAAAGames}.

Rendering and physics are key functionalities of video game
engines that are widely present in almost all video games.
Both Unity and Unreal Engine can be used to make games with simple or very complex physics. 
\textit{Kerbal Space Program} and \textit{Rocket League} 
are two examples of well-known games with complex physics made with Unity and Unreal Engine, respectively.
\textit{Kerbal Space Program} simulates real-world orbital mechanics and space travel physics. It includes elements like gravitational forces from planets and moons, orbital trajectories, the physics of thrust and propulsion in a vacuum, atmospheric drag, and re-entry heat. 
\textit{Rocket League} combines the physics of driving with the physics of a bouncing, spinning soccer ball. The game realistically simulates the momentum and inertia of both the cars and the ball and accurately simulates the interactions between the cars and the ball, including how the ball bounces and reacts to hits. All in all, comparable results of the key elements of visuals and physics can be achieved with both Unity and Unreal engine; however, the developers do not know the impact of the engines in terms of energy consumption because it has not been explored so far.




\section{Measuring the energy consumption of software}
\label{sec:MeasEner}

There are two different methodologies for this purpose: The first one resorts to specialized software tools that can estimate the energy consumed by the hardware when a given software is being used, such as PowerAPI or Joulemeter. These tools do not directly measure the energy consumed but rather estimate it from a model that is fed with parameters that the software collects while the object of study is being executed. These methods have the advantage that they are generally effortless, easy to apply, and produce valuable results at different levels of granularity. However, since they only estimate the energy consumed, the accuracy of the results depends on the fit of the model to the situation being analyzed.

The other methodologies available are those that use measuring equipment connected to a suitably instrumented computer to measure the energy consumed by the computer as a whole or by the different components to be studied while the software under study is running. This approach has the advantage over the previous one of producing real energy consumption results, whose accuracy depends only on the quality of the equipment and the measurement procedure. However, it requires instrumentation of the computer on which the test is to be performed and can be more costly and complicated to implement.

Having a physical device that allows us to measure the energy consumption of a computer running a certain software is not enough to guarantee that the results are sufficient to draw conclusions about it. It is necessary to define a complete process that guarantees the rigor and consistency of the studies so that the conclusions obtained are scientifically relevant. In this work, we use the process defined by Mancebo et al.~\cite{mancebo2021process}, the stages of which are described in Section~\ref{sec:ExpProc}.

\section{Experiment description}
\label{sec:ExpDescr}
The goal of this study is to assess the influence of the use of different video game engines on the energy consumption of a commercial video game. There are no public video games that have two identical versions programmed with Unity in one of them and Unreal Engine in the other for comparison. In practice, studios decide on the engine at the beginning of development and rarely change engines during development because it is a very costly decision.
Furthermore, if we were to take a complete video game, it would not be possible to isolate engine performance for key aspects such as visuals and physics. Video game scenes simultaneously combine multiple aspects such as physics simulations or mesh rendering, among others.
For example, in the video games shown in Figure~\ref{fig:indieAndAAAGames}, many of the video game engine`s functionalities are working simultaneously with all of the needed calculations derived from: physics simulations (e.g., when the character moves, jumps, or shoots physics simulations have to be performed); mesh rendering (e.g., every 3D visual element needs the rendering of its mesh with its textures); animation system operations (e.g., when the character moves, it needs to look like it is walking); and others.


Consequently, to compare video game engines, we have developed the same scenarios representing key aspects of the video games for both engines. Video games might differ one from another in interaction and mechanical aspects, but they require common aspects that include 2D and 3D rendering, physics, sound, and Artificial Intelligence~\cite{kasurinen2016games,musil2010survey,aleem2016game}. Currently, the most developed type of games are 3D games (47\% vs 36\% for 2D games~\cite{slashdata2022}). Three of the most common systems in 3D video games are physics, 3D rendering, and animation systems. That is why we decided to create three scenarios that focus on these systems in order to carry out the consumption comparison between Unity and Unreal Engine.

The first scenario (Physics scenario) was built to evaluate the physics aspects of the engine. 
The scenario instantiates 1000 cubes. Each cube has a mass of 50 kilograms, a size of 1 meter on each axis (X, Y, Z), and physics simulation and collision detection enabled. A random force of between 1 and 80 Newtons is applied to each cube on each frame, pushing them continuously toward the origin position. That will make the cubes collide with each other while they keep receiving the force toward the origin. Collisions of this nature are common in video games such as the popular Doom or the Battlefield series, where there are shots and explosions affecting players, non-player characters (NPCs), and other scene elements.
The second and third scenarios were built to evaluate the visual aspects of the engine
: Static Meshes rendering and animated Dynamic Meshes rendering.
Static meshes are 3D models that do not change shape or move during gameplay. They are often used for environment and background elements, such as buildings and furniture. Dynamic meshes are 3D models that can change shape and move in real time. They are often used for characters and interactive elements, such as NPCs and vehicles.

In the second scenario (implementing Static Meshes), 1000 static meshes are instantiated and stay in place, making the engine render them for the duration of the scenario.
In the third scenario (implementing Dynamic Meshes), 1000 dynamic meshes with a simple animation are instantiated and stay in place.
 The animation consists of the character stabbing with both hands, which is a common action in many fantasy video games, such as the popular video game \textit{Baldur's Gate III}.

The mesh and textures for both the second and third scenarios and the animation for only the third scenario have all been downloaded from \textit{Mixamo} (\url{https://www.mixamo.com/}). \textit{Mixamo}, which is owned by Adobe Inc., allows the downloading of ready-to-use professional 3D character meshes and animations for personal, commercial, and non-profit projects, including video games.

\begin{figure*}[h]
\centering
\includegraphics[width=.75\textwidth]{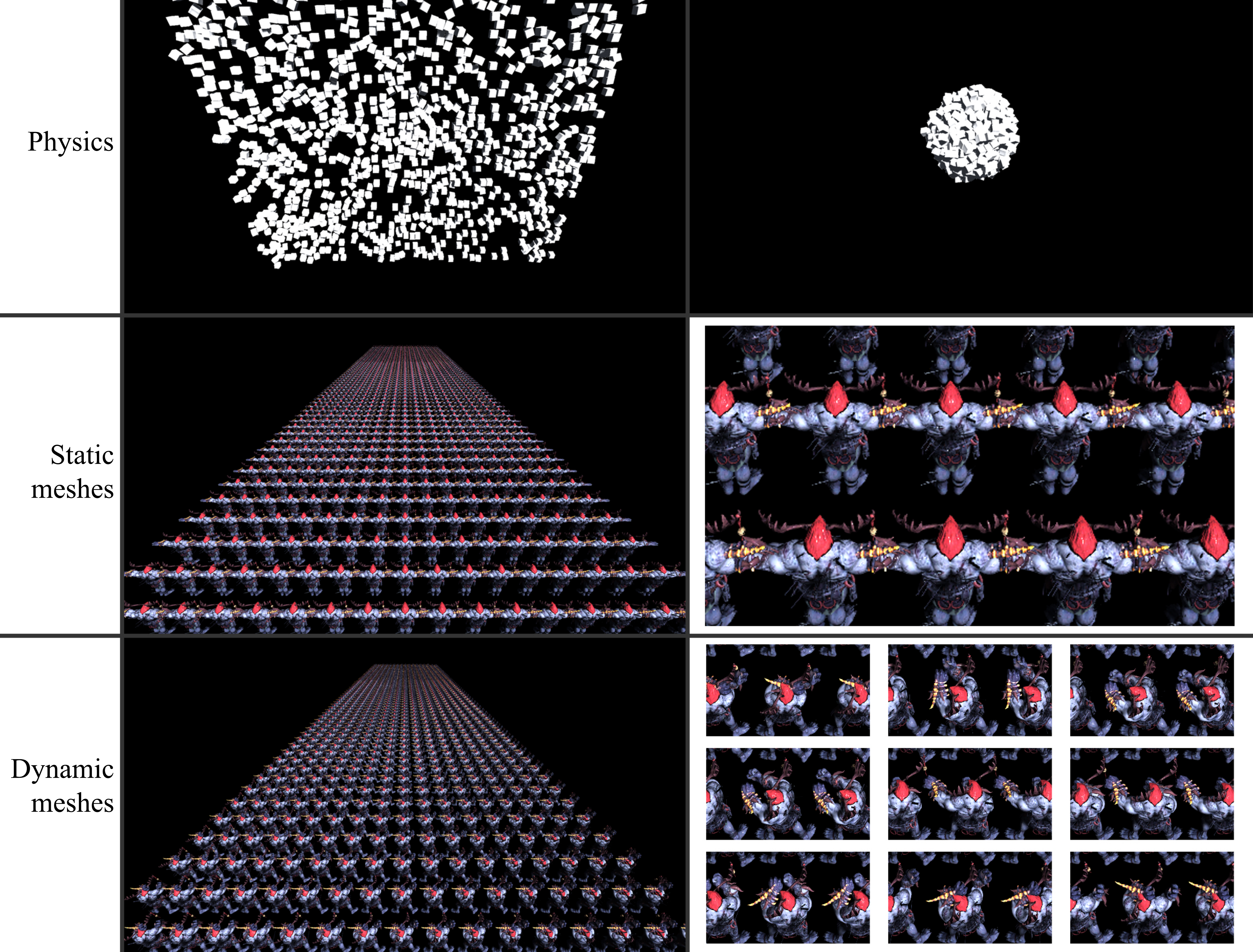}
\caption{The three scenarios implementing Physics, Static Mesh rendering, and Dynamic Mesh rendering.}
\label{fig:scenarios}
\end{figure*}

Images of the implementation of all three scenarios is shown in Figure~\ref{fig:scenarios}. Each row of the figure shows a different scenario. The first row contains two images from the execution of the Physics scenario, where 1000 cubes start scattered and then receive forces towards the center. The second row contains an image from the execution of the Static Meshes scenario, followed by a zoomed-in image of it so that the static mesh and its textures can be seen better. Lastly, the third row contains an image from the execution of the animated Dynamic Mesh scenario, followed by nine zoomed-in captures at different times of the execution which show the animation of the dynamic mesh.

The implementations of the scenarios in both engines have the same number of elements and complexity 
between the engines. 
The scenarios in both engines are built from an empty level where any skybox or atmosphere of any type has been removed, along with other elements that might have been in the default scene or level. The only elements in the scenarios are:
\begin{enumerate}
    \item One directional light with default configuration values except for the intensity, which has been modified to have the light in both engines with a similar intensity.
    \item The scenario element (i.e., an empty GameObject placed in the Scene in Unity or a Pawn instantiated in the PlayerStart on play in Unreal Engine) with default values and containing both the camera and a script that generates all of the objects (i.e., Prefab instances in Unity or Blueprint Actors instances in Unreal Engine) needed for each scenario (i.e., the cubes for the Physics scenario and the static and dynamic meshes for the other two scenarios).
    This script in Unity is programmed in C\# with a \textit{MonoBehaviour} class of Unity inheritor, as is expected to be done in this engine. In Unreal Engine, it is programmed with a combination of Blueprints visual scripting and C++ with an \textit{APawn} class of Unreal Engine inheritor, as is expected to be done in this engine.
\end{enumerate}

We used the latest available versions of both engines at the time when the different scenarios were implemented: 2022.3.6f1 for Unity and 5.2.1 for Unreal Engine. The scenarios were implemented by the second author, who happens to be a professional game developer himself, and have been validated as equivalent between the two engines by two different professionals in the industry (not involved in this paper).

All six applications, i.e., the versions of the three scenarios in the two engines, are available at the following URL for replication purposes: \url{https://svit.usj.es/unreal-vs-unity-energy-comparative-analysis/}

Since our aim is to compare the energy consumption of these two commercial game engines, the Research Questions for this work are built around this goal:
\begin{itemize}
    \item RQ1 – Is there a relationship between the energy consumption needed by a physics-related task at run time and the engine used during the development?
    \item RQ2 – Is there a relationship between the energy consumption needed by a static mesh rendering-related task at run time and the engine used during the development?
     \item RQ3 – Is there a relationship between the energy consumption needed by an animated dynamic mesh rendering-related task at run time and the engine used during the development?
\end{itemize}

\section{Experimental procedure}
\label{sec:ExpProc}

The process we follow in the execution of our experiment is defined by~\cite{mancebo2021process}. Based on a review of good practices in measuring software energy consumption, the process defined in~\cite{mancebo2021process} improves the reliability and consistency of the measurements, so that reproducibility and comparison of different studies is facilitated.

\subsection{Phases of the procedure}

\begin{figure}
\centering
\includegraphics[width=1\columnwidth]{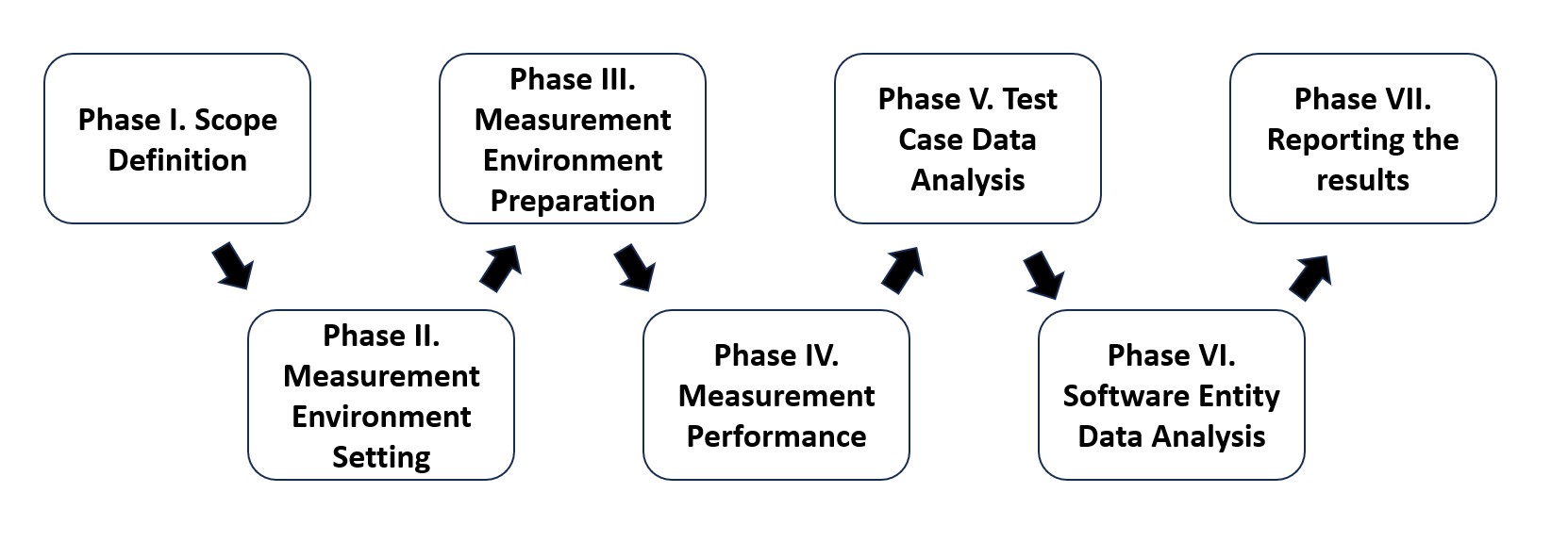}
\caption{Process for evaluating the energy efficiency of software}
\label{fig:diagramaGSMP}
\end{figure}

We followed the Green Software Measurement Process (GSMP), as proposed by Mancebo~et~al.~\cite{mancebo2021feetings}. This process consists of a series of phases outlined in Figure~\ref{fig:diagramaGSMP}, encompassing all the necessary steps for conducting a thorough analysis of software energy consumption during execution.

\subsubsection{Phase I. Scope Definition} 
\label{sec:scope}
During this stage, a complete set of requirements is obtained for evaluating energy efficiency. Furthermore, clear definitions of the software under study and the test cases to be analyzed are established.

Our study aims to compare the energy consumption during the execution of three representative scenarios of real video games, programmed with two different game engines. It is worth emphasizing that our objective is to find out if there is any significant difference in energy consumption between scenarios implemented with different engines. The goal is not to discover what this difference is exactly because this will always depend on the environment in which the applications are run.


Special care has to be taken to actively ensure that there are no background processes beyond those required by the operative system. The consumption of the processes required by Windows needs to be measured separately and discounted from the results as a baseline.

 
\subsubsection{Phase II. Measurement Environment Settings}
The purpose of this phase is the definition of the measurement environment that will be used in the software energy consumption assessment. 

The measurement environment of the Framework for Energy Efficiency Testing to Improve Environmental Goals of the Software (FEETINGS)~\cite{mancebo2021feetings} (to which GSMP also belongs) is used to evaluate the energy efficiency of the software. It is based on actual measurements using a device connected to a computer. Specifically, there are two elements for consumption measurement in FEETINGS: 

\begin{enumerate}
    \item The Efficient Energy Tester (EET), which is a device designed to measure the power requirement of a group of hardware components and the execution time of the Software Entity when running on the Device Under Test (DUT).
    \item The ELLIOT, which is the software application that processes and analyzes the data collected by the EET.
\end{enumerate}
The DUT is a desktop computer utilized for executing the test cases and conducting time and power measurements. In our experiment, the DUT had the specifications outlined in Table~\ref{tab:dutSpecs}.

\begin{table}[]
\resizebox{\columnwidth}{!}{%
\begin{tabular}{@{}lllll@{}}
\toprule
Monitor       & Falkon Q2702S 27” 2K                         &  &  &  \\ 
Motherboard   & Asus Prime B460-Plus                         &  &  &  \\
Processor     & i7 10700                                &  &  &  \\
RAM           & 4 modules of 32GB DDR4 Kingston 2666MHz CL16 &  &  &  \\
Graphics card & Zotac Gaming GForce RTX 3060 12 GB GDDR6     &  &  &  \\
Hard Disk     & Hard Disk \& Kingston SSD A400 – 480GB SATA  &  &  &  \\
Power supply  & Energy PS901SX 900W                          &  &  &  \\
O.S.          & Windows 11 Pro                               &  &  &  \\ \bottomrule
\end{tabular}%
}
\caption{DUT specifications}
\label{tab:dutSpecs}
\end{table}

The energy consumption of the processor, the HDD, and the overall energy consumption of the device are accepted measurements in green software studies~\cite{Calero21}.  However, in the case of video games, it is also necessary to measure the energy consumption of the graphics card as video games use it intensively.



The final power measurements are computed by subtracting the baseline power required (background processes before launching the scenarios) from the real measurements recorded during the experiment (after launching the scenarios).

\subsubsection{Phase III. Measurement Environment Preparation}
\label{sec:prep}
The primary objective of this phase is to prepare for the power measurements and configure the measurement environment accordingly.

EET has a power supply that must be connected to the device under test where the software is executed, replacing the power supply of the DUT.

 The power of the computer cooling system may influence the power required by the whole system if, in one of the scenarios, the computational demand causes the processor to overheat. To verify that this effect does not occur, two runs of the three scenarios with the two engines in different order have been performed to check if the order in which the different cases are run alters the results. The order of the two runs, labeled V1 and V2, is as follows:

\begin{itemize}
    \item 	V1: Unreal Engine-Dynamic Mesh, Unreal Engine-Static Mesh, Unreal Engine-Physics, Unity-Dynamic Mesh, Unity-Static Mesh, Unity-Physics
    \item V2: Unity-Physics, Unity-Static Mesh, Unity-Dynamic Mesh, Unreal Engine-Physics, Unreal Engine-Static Mesh, Unreal Engine-Dynamic Mesh
\end{itemize}

The power required for each scenario and the execution time are measured and recorded. Energy consumption can then be calculated by multiplying the average power by the execution time.  All of the scenarios were executed during the same time (60 seconds) for the different scenarios to be comparable in terms of energy consumption. 

The executable files of the six scenarios for the engines are copied to the DUT.
After checking that no other software is running in the background, the runs are then executed through a .BAT file. 
Both the sequence of executions and the number of repetitions have been detailed in the .BAT file.

To avoid undesirable effects and to guarantee the reliability of the analysis and the statistical results obtained, both runs (V1 and V2) were recorded 30 times. By repeating the measurement 30 times, it is possible to detect if there is any other software running in the background, since the results would then present differences between them. 
All of the power measurements that have been collected by the EET are saved onto a file for further processing.

\subsubsection{Phase IV. Perform the Measurements}
\begin{figure}[h]
\centering
\includegraphics[width=1\columnwidth]{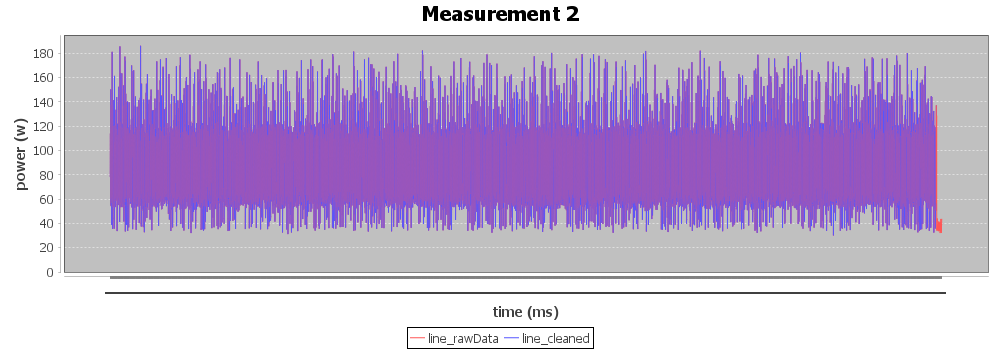}
\caption{Plot of the power measurements of one scenario as collected by the EET.}
\label{fig:raw}
\end{figure}

During this phase, power measurements are carried out and the raw power data taken from the measuring instrument is collected.  A sample plot of the raw data as collected by the EET is shown in Figure~\ref{fig:raw}. Approximately 6000 individual measurements are recorded, which corresponds to 60 seconds of recording at a sampling frequency of 100 Hz.

\subsubsection{Phase V. Test Case Data Analysis}
In this phase, the raw data of power obtained by the measuring instrument is processed using ELLIOT. As we mentioned previously, all of the measurements are repeated 30 times. Then, this raw dataset is cleaned by removing invalid measurements. A measurement is considered invalid if it corresponds to a wrong execution detected because all of the values obtained are not consistent with the rest of the results for the other executions, or it corresponds to an outlier, in which one or more values are either too high or too low compared to the rest of the results for the other executions. The number of executions removed from the exercise is shown in Table~\ref{tab:outliers}.

\begin{table}[h]
\resizebox{\columnwidth}{!}{%
\begin{tabular}{@{}lrrrrrr@{}}
\toprule
                       & \multicolumn{2}{c}{Dynamic Mesh} & \multicolumn{2}{c}{Static Mesh} & \multicolumn{2}{c}{Physics} \\ \midrule
                       & Unity       & UE      & Unity      & UE      & Unity    & UE    \\
Invalid   measurements & 4           & 7                  & 5          & 9                  & 4        & 9                \\
Valid   executions     & 26          & 23                 & 25         & 21                 & 26       & 21               \\ \bottomrule
\end{tabular}
}
\caption{Number of valid and invalid executions per scenario in Unity and UE (Unreal Engine)}
\label{tab:outliers}
\end{table}

 Once the data has been processed and prepared with ELLIOT, it also calculates the descriptive statistics of the values obtained. These statistics are shown in Table~\ref{tab:comp SM}, Table~\ref{tab:comp DM}, and Table~\ref{tab:comp PH}.

\begin{table*}[]
 \centering
\begin{tabular}{@{}
>{\columncolor[HTML]{FFFFFF}}l 
>{\columncolor[HTML]{FFFFFF}}l 
>{\columncolor[HTML]{FFFFFF}}l 
>{\columncolor[HTML]{FFFFFF}}l 
>{\columncolor[HTML]{FFFFFF}}l 
>{\columncolor[HTML]{FFFFFF}}l 
>{\columncolor[HTML]{FFFFFF}}l 
>{\columncolor[HTML]{FFFFFF}}l 
>{\columncolor[HTML]{FFFFFF}}l @{}}
\toprule
Average   without baseline (W)           & \multicolumn{2}{c}{\cellcolor[HTML]{FFFFFF}HDD} & \multicolumn{2}{c}{\cellcolor[HTML]{FFFFFF}Graphics Card} & \multicolumn{2}{c}{\cellcolor[HTML]{FFFFFF}Processor} & \multicolumn{2}{c}{\cellcolor[HTML]{FFFFFF}DUT} \\ \midrule
Video Game Engine                                & UE                  & Unity                 & UE                       & Unity                      & UE                    & Unity                     & UE                 & Unity                  \\
Mean                                     & 1,95                    & 1,98                  & 104,53                       & 93,20                      & 22,75                     & 19,26                     & 422,51                 & 377,18                 \\
Mean StandardDeviation (W)   (Time in s) & 1,27                    & 1,09                  & 38,21                        & 37,11                      & 15,80                     & 18,42                     & 33,55                  & 20,74                  \\
Mean Median (W) (Time in s)              & 1,65                    & 1,84                  & 102,71                       & 71,08                      & 17,61                     & 11,92                     & 426,31                 & 378,55                 \\
Min Record (W)                           & 0,03                    & 0,03                  & 30,94                        & 30,68                      & 0,3                       & 0,3                       & 166,277                & 235,15                 \\ 
Max Record (W)                           & 8,16                    & 7,47                  & 194,44                       & 193,5                      & 147,22                    & 144,25                    & 490,963                & 457,789                \\ \bottomrule
\end{tabular}
\caption{Comparison between Unity and UE (Unreal Engine). Static Mesh scenario}
\label{tab:comp SM}
\end{table*}

\begin{table*}[]
\centering
\begin{tabular}{@{}
>{\columncolor[HTML]{FFFFFF}}l 
>{\columncolor[HTML]{FFFFFF}}l 
>{\columncolor[HTML]{FFFFFF}}l 
>{\columncolor[HTML]{FFFFFF}}l 
>{\columncolor[HTML]{FFFFFF}}l 
>{\columncolor[HTML]{FFFFFF}}l 
>{\columncolor[HTML]{FFFFFF}}l 
>{\columncolor[HTML]{FFFFFF}}l 
>{\columncolor[HTML]{FFFFFF}}l @{}}
\toprule
Average without baseline (W)             & \multicolumn{2}{c}{\cellcolor[HTML]{FFFFFF}HDD} & \multicolumn{2}{c}{\cellcolor[HTML]{FFFFFF}Graphics Card} & \multicolumn{2}{c}{\cellcolor[HTML]{FFFFFF}Processor} & \multicolumn{2}{c}{\cellcolor[HTML]{FFFFFF}DUT} \\ \midrule
Video Game Engine                                & UE                  & Unity                 & UE                      & Unity                       & UE                    & Unity                     & UE                 & Unity                  \\
Mean                                     & 1,87                    & 1,87                  & 67,46                       & 86,45                       & 24,80                     & 29,53                     & 218,74                 & 296,59                 \\
Mean   StandardDeviation (W) (Time in s) & 1,28                    & 1,16                  & 22,88                       & 38,33                       & 33,50                     & 30,17                     & 21,26                  & 24,14                  \\
Mean   Median (W) (Time in s)            & 1,54                    & 1,71                  & 64,53                       & 65,60                       & 11,26                     & 16,98                     & 327,57                 & 404,98                 \\
Min Record (W)                           & 0,03                    & 0,03                  & 28,90                       & 28,15                       & 0,30                      & 0,30                      & 168,24                 & 252,97                 \\ 
Max Record (W)                           & 7,22                    & 7,22                  & 188,37                      & 190,72                      & 157,32                    & 154,35                    & 397,96                 & 499,35                 \\ \bottomrule
\end{tabular}
\caption{Comparison between Unity and UE (Unreal Engine). Dynamic Mesh scenario}
\label{tab:comp DM}
\end{table*}

\begin{table*}[]
\centering
\begin{tabular}{@{}
>{\columncolor[HTML]{FFFFFF}}l 
>{\columncolor[HTML]{FFFFFF}}l 
>{\columncolor[HTML]{FFFFFF}}l 
>{\columncolor[HTML]{FFFFFF}}l 
>{\columncolor[HTML]{FFFFFF}}l 
>{\columncolor[HTML]{FFFFFF}}l 
>{\columncolor[HTML]{FFFFFF}}l 
>{\columncolor[HTML]{FFFFFF}}l 
>{\columncolor[HTML]{FFFFFF}}l @{}}
\toprule
Average without baseline (W)           & \multicolumn{2}{c}{\cellcolor[HTML]{FFFFFF}HDD} & \multicolumn{2}{c}{\cellcolor[HTML]{FFFFFF}Graphics Card} & \multicolumn{2}{c}{\cellcolor[HTML]{FFFFFF}Processor} & \multicolumn{2}{c}{\cellcolor[HTML]{FFFFFF}DUT} \\ \midrule
Video Game Engine                               & UE                  & Unity                 & UE                       & Unity                      & UE                    & Unity                     & UE                 & Unity                  \\
Mean                                     & 1,91                    & 1,37                  & 62,39                        & 21,28                      & 63,49                     & 18,80                     & 413,58                 & 175,32                 \\
Mean StandardDeviation (W)   (Time in s) & 1,23                    & 0,81                  & 28,78                        & 4,63                       & 29,38                     & 22,78                     & 30,65                  & 9,39                   \\
Mean Median (W) (Time in s)              & 1,65                    & 1,26                  & 48,08                        & 21,07                      & 58,17                     & 9,22                      & 417,15                 & 174,97                 \\
Min Record (W)                           & 0,03                    & 0,03                  & 27,52                        & 2,87                       & 0,41                      & 0,30                      & 165,70                 & 125,15                 \\ 
Max Record (W)                           & 7,03                    & 6,88                  & 179,11                       & 63,99                      & 178,15                    & 142,46                    & 469,77                 & 244,23                 \\ \bottomrule
\end{tabular}
\caption{Comparison between Unity and UE (Unreal Engine). Physics scenario}
\label{tab:comp PH}
\end{table*}

\begin{figure*}
\centering
\includegraphics[scale=0.6]{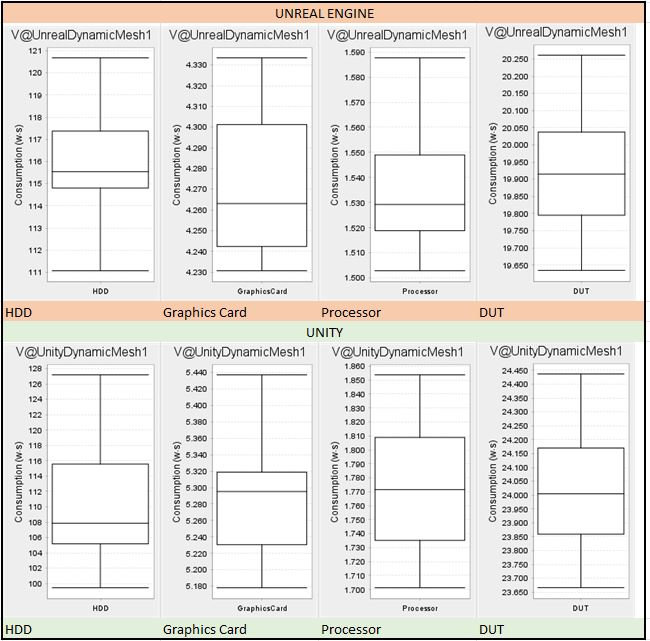}
\caption{Energy consumption boxplots for Dynamic Mesh scenario}
\label{fig:grafica DM}
\end{figure*}

\begin{figure*}
\centering
\includegraphics[scale=0.6]{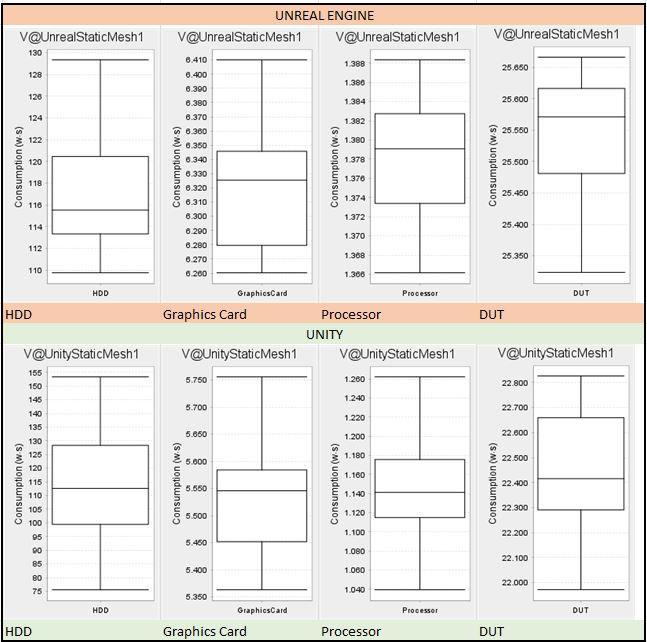}
\caption{Energy consumption boxplots for Static Mesh scenario}
\label{fig:grafica sm}
\end{figure*}

\begin{figure*}
\centering
\includegraphics[scale=0.6]{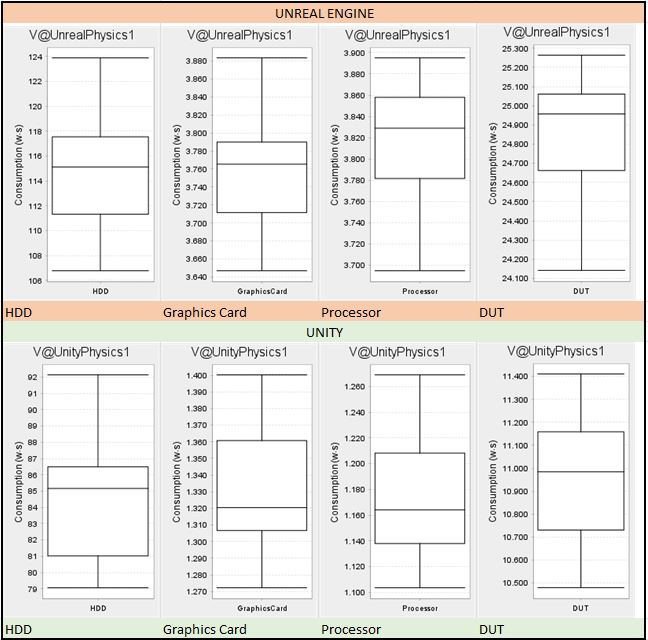}
\caption{Energy consumption boxplots for Physics scenario}
\label{fig:grafica ph}
\end{figure*}

Boxplots enable the examination of dataset distribution by dividing it into four equal parts, each containing an equal number of measurements. These parts are delineated by lines known as quartiles, with the groups being termed quartile groups.
In our case, the raw dataset is made up of 30 measurements, before deleting the invalid measurements. Energy consumption is calculated for each measurement by multiplying the average power by the execution time. The boxplots of the consumption data for our study are shown in Figure~\ref{fig:grafica DM}, Figure~\ref{fig:grafica sm}, and Figure~\ref{fig:grafica ph}.

Uneven sizes among the four sections of a boxplot indicate variability in measurements across different parts of the scale, with some sections showing similar values while others demonstrate greater variability.
In our case, if we look at the boxplots of each of the three scenarios comparing those run with Unity and Unreal Engine, we can see that there are no outstanding asymmetries. Boxplots are skewed left for the Dynamic Mesh scenario in Unity (Graphics Card), for the Static Mesh scenario (Graphics Card), and for the Physics scenario with Unity (HDD)  and Unreal Engine (DUT). Boxplots are skewed right for the Dynamic Mesh scenario (HDD), for the Dynamic Mesh scenario with Unreal Engine (HDD), and for the Physics scenario with Unity (Graphics Card and Processor). The rest of the scenarios show relatively symmetrical distributions.

If we compare the boxplots for each of the video game engines for the same component and scenario, it can observed that the median line lies outside the box of the comparison boxplot for all of the simulations except for the comparison between HDD consumption with Unity versus Unreal Engine in the Static Mesh scenario. This means that there are significant differences between the consumption of the scenarios executed with between Unity and Unreal Engine in almost all of the situations.

Lastly, we can explore the interquartile range to assess the dispersion of the data. As depicted in Figures~\ref{fig:grafica DM}, ~\ref{fig:grafica sm}, and ~\ref{fig:grafica ph}, overall, the length of the box is small, indicating limited data dispersion.

\subsubsection{Phase VI. Software Entity Data Analysis}
In this phase, the power required by the DUT during the executions of the different scenarios is presented, comparing the values of the scenarios executed with Unity and those executed with the Unreal Engine.

\begin{figure}
\centering
\includegraphics[width=1\columnwidth]{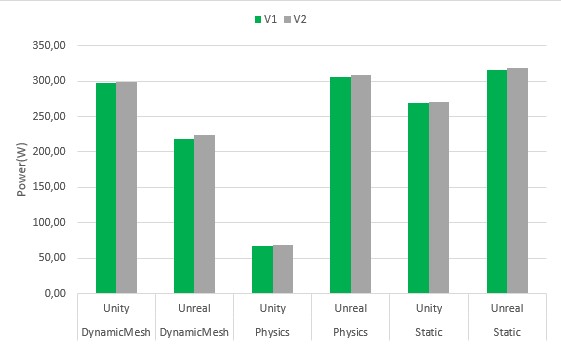}
\caption{DUT power required for sequences v1 and v2}
\label{fig:V1V2}
\end{figure}

Figure~\ref{fig:V1V2} shows the comparison between the power in each of the three scenarios for both engines of the exercise run according to the sequences described in Section~\ref{sec:prep}. In other words, sequences V1 and V2 run the same cases but in reverse order. This exercise aims to determine whether the order of execution of the different cases has any effect on the power required by the device for each of the scenarios. As shown in the graph, the differences are very small (between 1\% and 3\%), so we consider that the order in which the cases are executed does not alter the representativeness of the comparison between the Unity and Unreal Engine versions of the three scenarios.

We now move on to analyze the power required of the computer running each of the three scenarios with the two engines that constitute the main goal of the work. Table~\ref{tab:power SM}, Table~\ref{tab:power DM}, and Table~\ref{tab:power PH} show the mean power values obtained in each of the three scenarios of both engines. The columns show the power required by the hard disk, the graphics card, and the processor, and the last column reflects the power of the DUT 
as a whole.

Table~\ref{tab:power SM} shows the results of the comparison for the Static Mesh scenario. In this case, Unreal Engine is approximately 17\% less efficient in terms of the power required by the DUT than Unity. If we look at the power required by the different components, the difference is more than 18\% in processor power and more than 12\% in the case of the graphics card. There are no appreciable differences in hard disk data. 

Unreal Engine is known for its high-fidelity graphics and advanced rendering features. These results might be because these features can be more power-intensive and 
might favor higher visual quality at the cost of increased energy consumption. It is important to point out that the Static Mesh scenario in Unreal Engine and the one in Unity have the same visuals, that is, the very same mesh. Therefore,  the rendering of Unreal Engine comes at an energy cost even though there is no visual benefit.


Table~\ref{tab:power DM} shows the comparison between Unity and Unreal Engine for the Dynamic Mesh scenario.  In this scenario, the results of the previous scenario are reversed and Unreal Engine is more energy efficient than Unity, with a difference of more than 26\% in the power required by the DUT. If we look at the power required by the different components, the difference is 16\% in processor power and almost 22\% in the case of the graphics card. There is also no appreciable difference in hard disk data for this scenario.

These results denote that the animation of Unity might be less optimized for handling complex animations compared to the one used by Unreal Engine. In addition, may be that Unreal Engine might have rendering optimizations that specifically benefit dynamic meshes, since the consumption in the Static Mesh scenario is also greater than in the Dynamic Mesh scenario for this engine.


The Physics scenario comparison, shown in Table~\ref{tab:power PH}, is the one that shows the greatest percentage differences in power between the Unity and Unreal Engine executions. In this case, there are significant differences in hard disk power, which in the previous cases had not shown any discrepancy. The version developed with Unity requires 41\% less power than the version developed with Unreal Engine. This significant difference is amplified when the consumption of the graphics card and processor is analyzed, where the power requirement of the Unreal Engine version is more than three times bigger than the power requirement of the Unity version (218\% and 241\%, respectively), and 4.5 times bigger (351\%) when considering the power required by the complete DUT.

The origin of this difference in consumption might be in the physics engine employed by default in each of the game engines: Unity uses \textit{NVIDIA PhysX}, while Unreal Engine uses \textit{Chaos}. To date, there has been no study on the energy efficiency of these physics engines.

\begin{figure}
\centering
\includegraphics[width=1\columnwidth]{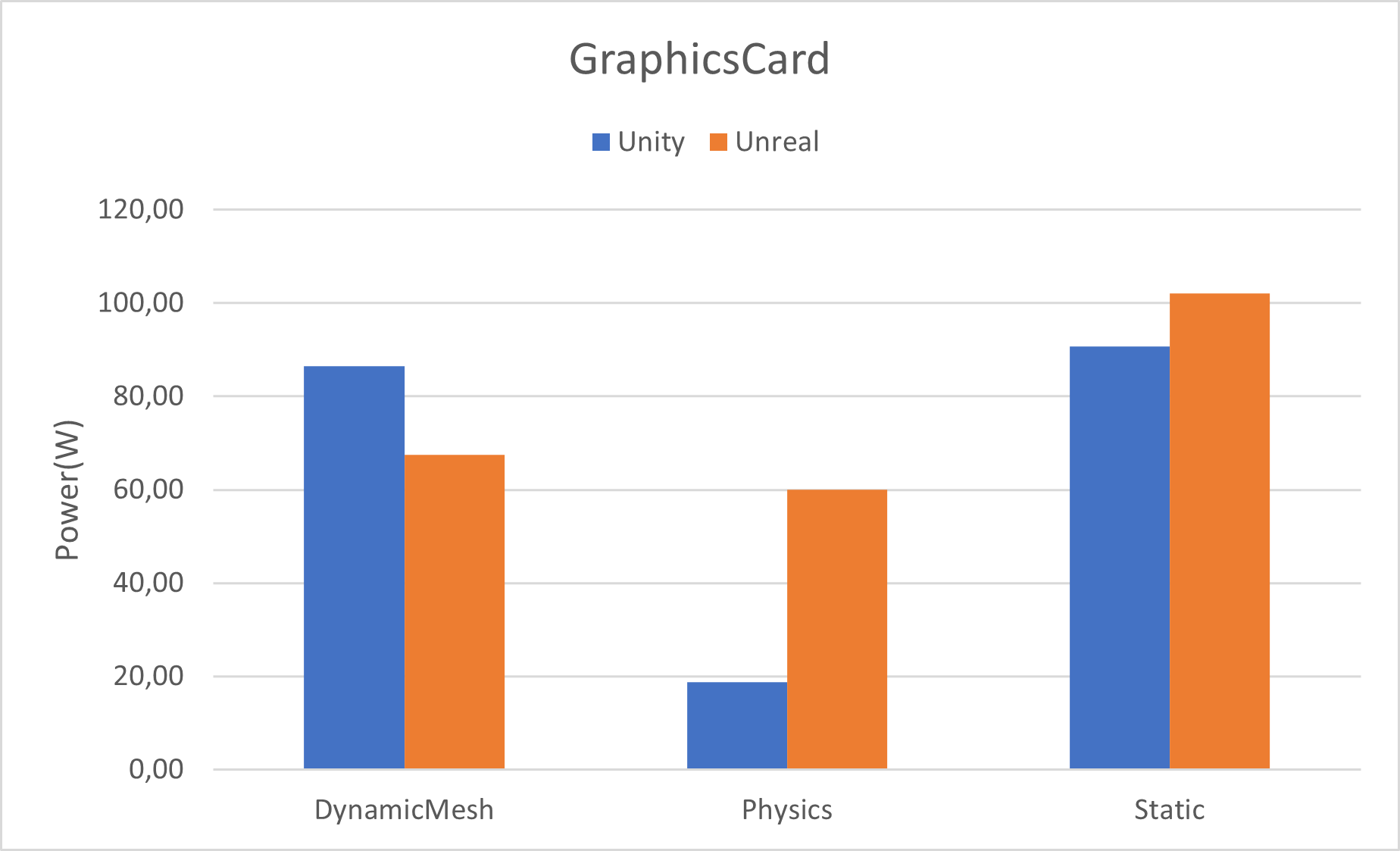}
\caption{Graphic card power required for the three scenarios}
\label{fig:GC}
\end{figure}

\begin{figure}
\centering
\includegraphics[width=1\columnwidth]{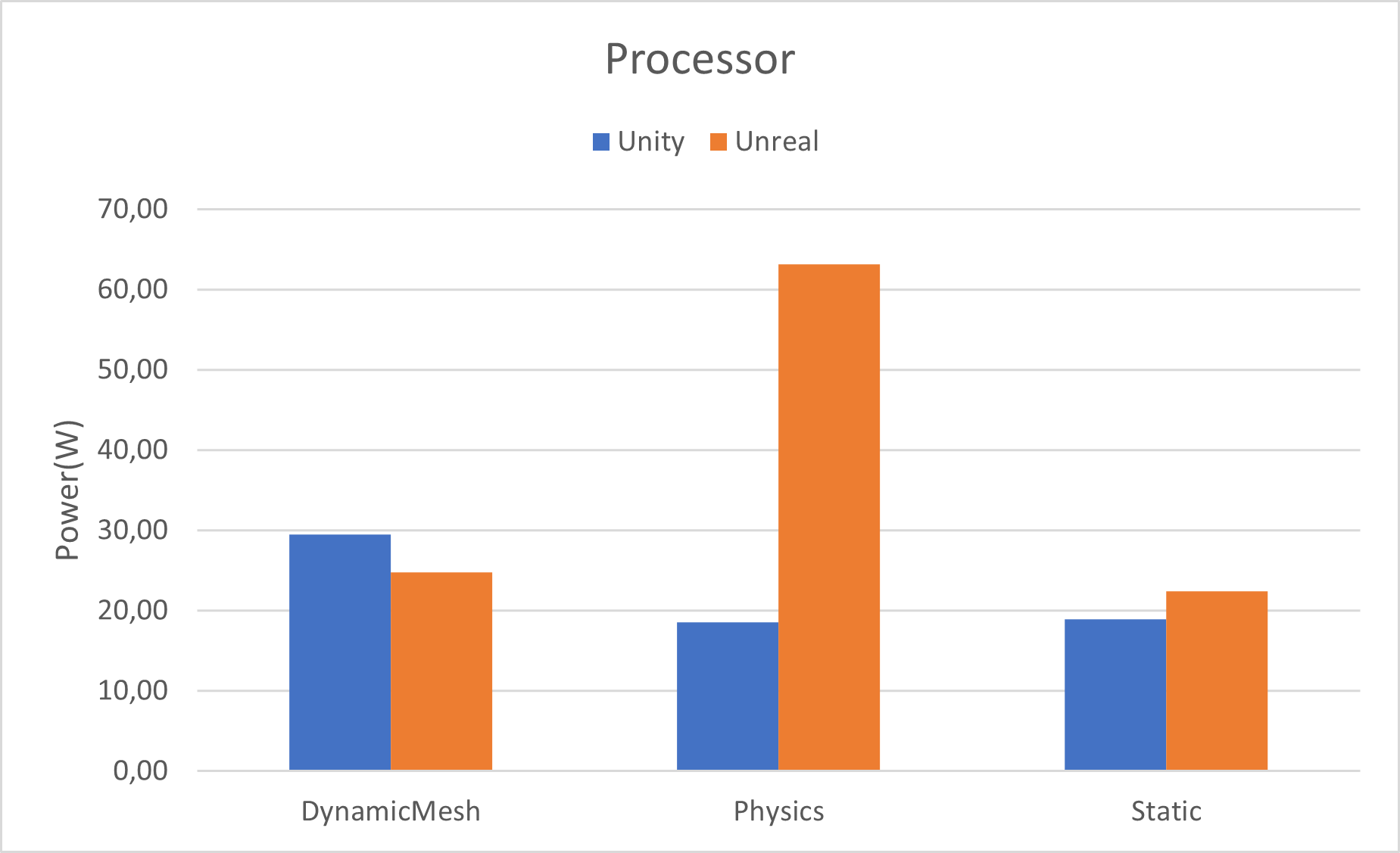}
\caption{Processor power required for the three scenarios}
\label{fig:pro}
\end{figure}

\begin{figure}
\centering
\includegraphics[width=1\columnwidth]{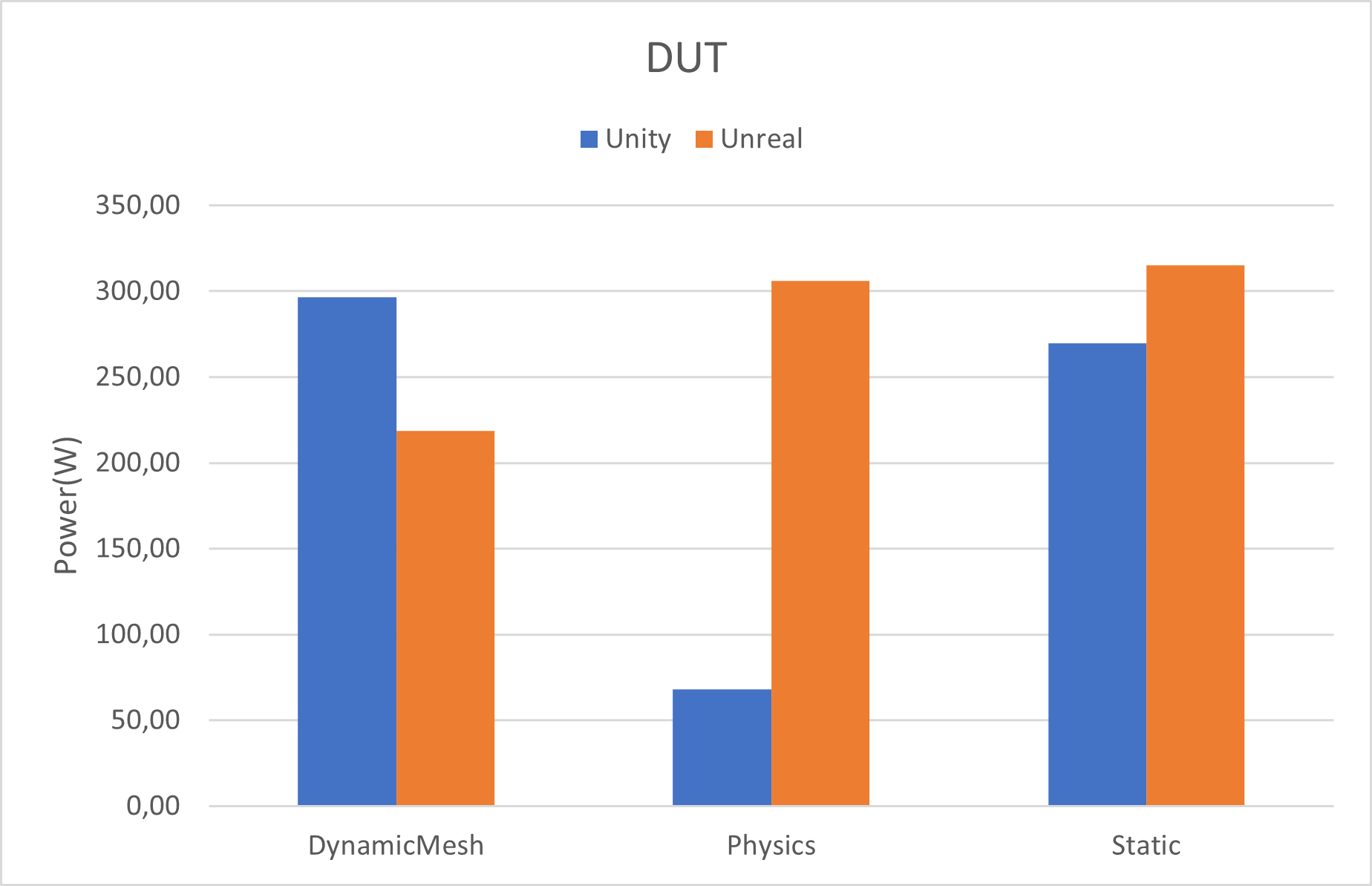}
\caption{DUT power required for the three scenarios}
\label{fig:DUT}
\end{figure}

Finally, we compare the power required by each of the different elements of the computer in the different scenarios, The power required by the hard disk is almost two orders of magnitude lower than that of the other elements included in the study (graphics card and processor) with small differences between the two engines. The power of the graphics card, (shown in Figure~\ref{fig:GC}); and the processor, (shown in Figure~\ref{fig:pro}), are comparable (same order of magnitude). The graphics card requires more power in the two rendering cases (Static and Dynamic) while the processor should be dominant in the case of Physics. This is so in the case of Unreal Engine, but in the case of Unity, the power required by the processor for the three cases is quite similar. The power required by the complete DUT for the three scenarios in both engines is also shown in Figure~\ref{fig:DUT}.

\begin{table}[]
\resizebox{\columnwidth}{!}{%
\begin{tabular}{@{}
>{\columncolor[HTML]{FFFFFF}}c 
>{\columncolor[HTML]{FFFFFF}}c 
>{\columncolor[HTML]{FFFFFF}}c 
>{\columncolor[HTML]{FFFFFF}}c 
>{\columncolor[HTML]{FFFFFF}}c 
>{\columncolor[HTML]{FFFFFF}}c @{}}
\toprule
Test case                             & Engine                           & HDD    & GraphicsCard & Processor & DUT      \\ \midrule
StaticMesh  (W)                          & Unity                            & 1,95   & 90,74        & 18,96     & 269,78   \\
StaticMesh (W)                            & UE                           & 1,92   & 102,07       & 22,45     & 315,11   \\
\multicolumn{2}{c}{\cellcolor[HTML]{FFFFFF}Power diff (Unity-UE)} & 1,36\% & -12,49\%     & -18,41\%  & -16,80\% \\ \bottomrule
\end{tabular}%
}
\caption{Power required (W) for Unity and UE (Unreal Engine). Static Mesh}
\label{tab:power SM}
\end{table}

\begin{table}[]
\resizebox{\columnwidth}{!}{%
\begin{tabular}{@{}
>{\columncolor[HTML]{FFFFFF}}c 
>{\columncolor[HTML]{FFFFFF}}c 
>{\columncolor[HTML]{FFFFFF}}c 
>{\columncolor[HTML]{FFFFFF}}c 
>{\columncolor[HTML]{FFFFFF}}c 
>{\columncolor[HTML]{FFFFFF}}c @{}}
\toprule
Test case                             & Engine                           & HDD     & GraphicsCard & Processor & DUT     \\ \midrule
DynamicMesh (W)                          & Unity                            & 1,87 & 86,45        & 29,53     & 296,59  \\
DynamicMesh (W)                           & UE                           & 1,87    & 67,46        & 24,80     & 218,74  \\
\multicolumn{2}{c}{\cellcolor[HTML]{FFFFFF}Power diff (Unity-UE)} & 0,01\%  & 21,97\%      & 16,02\%   & 26,25\% \\ \bottomrule
\end{tabular}%
}
\caption{Power required (W) for Unity and UE (Unreal Engine). Dynamic Mesh}
\label{tab:power DM}
\end{table}

\begin{table}[]
\resizebox{\columnwidth}{!}{%
\begin{tabular}{@{}
>{\columncolor[HTML]{FFFFFF}}c 
>{\columncolor[HTML]{FFFFFF}}c 
>{\columncolor[HTML]{FFFFFF}}c 
>{\columncolor[HTML]{FFFFFF}}c 
>{\columncolor[HTML]{FFFFFF}}c 
>{\columncolor[HTML]{FFFFFF}}c @{}}
\toprule
Test case                            & Engine                            & HDD      & GraphicsCard & Processor & DUT       \\ \midrule
Physics (W)                              & Unity                             & 1,34     & 18,82        & 18,50     & 67,92     \\
Physics (W)                              & UE                            & 1,88     & 59,93        & 63,19     & 306,17    \\
\multicolumn{2}{c}{\cellcolor[HTML]{FFFFFF}Power diff (Unity-UE)} & -40,92\% & -218,45\%    & -241,49\% & -350,78\% \\ \bottomrule
\end{tabular}%
}
\caption{Power required (W) for Unity and UE (Unreal Engine). Physics}
\label{tab:power PH}
\end{table}

\subsection{Answering the Research Questions}
We can now answer the research questions that were posed at the beginning of the article. Although all of the graphs are presented in power, we can draw conclusions about energy consumption since the run time was the same in all of the  scenarios (60 seconds).

\subsubsection{Answering RQ1 - Physics}
When asked whether there is a relationship between the energy consumption of the physics-related scenario between the versions coded in the two engines (RQ1), it is observed that there is a significant difference in consumption, with the  DUT consumption of the Unreal Engine version being almost 4,5 times higher than that of the Unity version (351 \%). This difference in the overall consumption of the device also appears in the consumption of the different components, (218 \% for the graphics card and 241 \% for the processor). For the hard disk consumption, the difference between the two engines is smaller (about 41\%), but significant. 

\subsubsection{Answering RQ2 - Static Mesh rendering}
When asked if there is a relationship between the energy consumption of the DUT  in a static mesh rendering task and the game engine used (RQ2), the answer is affirmative. There is a difference of 17\% in favour of the Unity version in the overall consumption of the device in this type of task. Once again, the differences in hard disk consumption are very small,(less than 2\%), while the differences in this case for both the consumption of the graphics card and the processor show differences ranging from 12\% to 18\%.

\subsubsection{Answering RQ3 - Animated Dynamic Mesh rendering}
Finally, for the third research question about whether there are differences in consumption between the versions coded in Unity and Unreal engine in the rendering of dynamic meshes, the results indicate that, in this case, it is the Unreal Engine that has 26\% lower energy consumption than the Unity engine. This difference in consumption is evident in the consumption of the graphics card, (22\% lower in the case of Unreal Engine and 16\% lower in the processor), while the consumption of the hard disk is virtually identical between the two versions.


\subsection{Putting the results in context} 

We can state that, in light of the results obtained, there are significant differences between the versions using one or the other video game engine in the case of the scenarios analyzed. Our research has confirmed significant differences in the
energy consumption of video game engines: 351\% in Physics
in favor of Unity, 17\% in Static Mesh in favor of Unity, and
26\% in Dynamic Mesh in favor of Unreal Engine.



By combining both engines and theoretically choosing the most efficient engine technology for each scenario (assuming the
simplification that the three components have a similar weight
in the context of a video game), it would be possible to outperform any of the two engines, as presented in Table~\ref{tab:optpower}. When comparing this data with a weighted average accounting for the market share of the video game engines, a global power saving (from the current situation) of 22\% can be estimated as shown in Table~\ref{tab:savings}.

\begin{table}[]
\resizebox{\columnwidth}{!}{%
\begin{tabular}{@{}rcrrr@{}}
\toprule
\rowcolor[HTML]{FFFFFF} 
\multicolumn{1}{l}{\cellcolor[HTML]{FFFFFF}\textbf{}} & \multicolumn{4}{c}{\cellcolor[HTML]{FFFFFF}Power measurements   DUT  ( W)}                                                                                                                                                     \\ \midrule
\rowcolor[HTML]{FFFFFF} 
\multicolumn{1}{l}{\cellcolor[HTML]{FFFFFF}}          & Static Mesh                                       & \multicolumn{1}{c}{\cellcolor[HTML]{FFFFFF}Dynamic Mesh} & \multicolumn{1}{c}{\cellcolor[HTML]{FFFFFF}Physics} & \multicolumn{1}{c}{\cellcolor[HTML]{FFFFFF}Average power} \\
\rowcolor[HTML]{FFFFFF} 
\textbf{Unity}                                                 & \multicolumn{1}{r}{\cellcolor[HTML]{A9D08E}269,8} & 296,6                                                    & \cellcolor[HTML]{A9D08E}67,9                        & 211,4                                                     \\
\rowcolor[HTML]{FFFFFF} 
\textbf{Unreal Engine}                                                & \multicolumn{1}{r}{\cellcolor[HTML]{FFFFFF}315,1} & \cellcolor[HTML]{A9D08E}218,7                            & 306,2                                               & 280,0                                                     \\
Best case                                             & \multicolumn{1}{l}{269,8}                         & \multicolumn{1}{l}{218,7}                                & \multicolumn{1}{l}{67,9}                            & \cellcolor[HTML]{FFFFFF}185,5                             \\ \bottomrule
\end{tabular}%
}
\caption{ Power estimation. Average of scenarios. The optimum per scenario is highlighted}
\label{tab:optpower}
\end{table}

According to the worldwide consumption figures presented in Section \ref{sec:Introduction} 22\% represents a potential saving of between 51 TWh and 76 TWh per year, which is equivalent to the annual consumption of at least 13 million European households.
There is so much opportunity for improvement because of the huge popularity of video games. More than 40\% of the world's population plays games, and, in recent years, this  has grown at a rate of 15\% per year. Moreover, if in the future teleworking leads to virtual environments made with video game engines, these numbers will grow even more. We expect these results to raise awareness regarding the energy consumption of video game engines and to motivate a new branch of research in that direction.




\begin{table}[]
\resizebox{\columnwidth}{!}{%
\begin{tabular}{lllr}
\hline
\rowcolor[HTML]{FFFFFF} 
\multicolumn{1}{|c}{\cellcolor[HTML]{FFFFFF}\textbf{Engine}} & \multicolumn{1}{c}{\cellcolor[HTML]{FFFFFF}Market share} & \multicolumn{1}{c}{\cellcolor[HTML]{FFFFFF}Average power (W)} & \multicolumn{1}{c}{\cellcolor[HTML]{FFFFFF}Optimum power (W)} \\ \hline
\rowcolor[HTML]{FFFFFF} 
Unity                                                        & 0,38                                                     & 211,4                                                         & 185,5                                                         \\
\rowcolor[HTML]{FFFFFF} 
Unreal Engine                                                & \cellcolor[HTML]{FFFFFF}0,15                             & 280,0                                                         & \cellcolor[HTML]{FFFFFF}185,5                                 \\
\rowcolor[HTML]{FFFFFF} 
Others                                                       & 0,47                                                     & \cellcolor[HTML]{FFFFFF}245,7                                & 185,5                                                         \\ \hline
\multicolumn{1}{r}{Average  per user(W)}                     &                                                          & 237,8                                                         & 185,5                                                         \\
\multicolumn{1}{r}{}                                         & \multicolumn{1}{r}{}                                     & \multicolumn{1}{r}{Potential savings}                            & \multicolumn{1}{l}{-22,0\%}                                  
\end{tabular}%
}
\caption{Global estimated power savings (The average of the measured power of  Unity and Unreal Engine is used as an estimation for the "Others" category).}
\label{tab:savings}
\end{table}

\section{Threats to validity}
\label{sec:threatsToValidity}

In this section, we use the classification of threats to validity of Wohlin~et~al.~\cite{VAL:Wohlin12}.

\textbf{Construct validity:} This aspect of validity reflects  "the degree to which the independent and the dependent variables are accurately measured by the measurement instruments used in the experiment”.

\begin{itemize}
\item{Dependent variables.}
In this experiment, the dependent variables are the consumption of energy required for executing the different test scenarios. These consumptions were objectively measured using the Efficient Energy Tester, which has been validated as a dependable device for assessing the energy efficiency of software execution~\cite{Calero21}. Additionally, each execution was repeated 30 times, and all scenarios were executed on the same physical machine, operating system, and configuration.
\item{Independent variables.}
The independent variables are the use of Unity or Unreal Engine as the video game engine of the application and the selection of the three scenarios to represent the consumption behavior of a typical video game.
 
\end{itemize}

\textbf{Internal Validity:} This aspect of validity becomes crucial when exploring causal relationships. There is a potential risk that the factor under investigation might be influenced by other unaccounted variables.

Due to the nature of both experiments, all of the variables were controlled, thereby minimizing potential threats to internal validity. Specifically, the test cases are the same and run the same scenarios in both video game engines (Unity and Unreal Engine). It is worth noting again that the repetition of the experiments 30 times during the same time also helped to minimize any possible bias. In addition, subtracting the baseline power in the measurements and removing invalid measurements (wrong executions or outliers) also helped to ensure that the dataset was valid for analyzing and answering the Research Questions.

\textbf{External Validity:}
 This aspect of validity focuses on the extent to which findings can be generalized and are relevant to other cases. Enhanced external validity indicates a greater ability to apply the results of an empirical study to real-world software engineering practices.

To minimize the risk of the Unity and Unreal Engine implementations not being comparable, the scenarios have been validated as being equivalent between the two engines by two different professionals in the industry. Furthermore, the scenarios are representative of key aspects of video games (visual and physical) that are present in the vast majority of video games being marketed today.

Another threat to the external validity is the fact that the measurements are obtained for a specific computer (the DUT being used). However, we believe that using an already validated process and measurement device makes the results more reliable because, although the actual consumption data may be different on another computer, the relative differences between those values would be similar to those obtained in this case. The study was carried out with a computer that can be considered "standard" for gaming, although there could be more significant changes if a particularly powerful computer were chosen in some aspect.


\section{Conclusions}
\label{sec:Conclusion}

The video games sector has experienced strong growth in recent years, both in terms of the number of users and the complexity of the games. This necessarily implies a significant increase in the energy consumed by this activity.
However, the environmental awareness of this sector, particularly the efforts to reduce the energy consumption of the billions of gamers, is still far from maturity, and the magnitude of the energy consumed makes this a relevant issue for the research community.

Within the set of tools available to developers, game engines provide game developers with essential tools and functionalities, enabling them to focus on the creative and unique aspects of the game rather than having to program all of the basic functionalities from scratch.

The influence of the game engine on the energy consumption of the application has been studied in this work. We selected the most popular video game engines (Unity and Unreal Engine) and three representative scenarios of common tasks that are relevant to the majority of video games (Static Mesh rendering, Animated Dynamic Mesh rendering, and Physics simulations). The power required for the different implementations has been carefully measured in a controlled environment.  Since the execution time was the same for all of the scenarios, the power measurements and energy consumption are strictly proportional. 

The results obtained show that there are clear differences in energy consumption in the different scenarios for the two engines. These differences can be summarized as follows:

\begin{mdframed}
\textit{Unreal Engine has a lower consumption than Unity in the Dynamic Mesh rendering scenario (17\%).} \\\\
\textit{Unity has a lower consumption than Unreal Engine in the Static Mesh scenario (26\%).} \\\\
\textit{Unity has a lower consumption than Unreal Engine in the Physics simulation scenario (351\%).}
\end{mdframed}

The differences are significant, both in the consumption of individual components such as the graphics card or the processor and in the overall power of the computer. Therefore the selection of an energy-efficient engine for the tasks included in the game could have a direct impact on the overall consumption. 

The fact that Unity and Unreal Engine outperform each other in energy efficiency based on the 
selected scenario, indicates that there is potential interest in finding the most energy-efficient video game engines, identifying best practices, and implementing them in a more refined, energy-efficient video game engine technology. 

There are still relevant functionalities implemented in both engines that are common in many different video games that should be analyzed, such as the sound system and the navigation system for AI path-finding.
Also, one of the reasons for choosing the usage of commercial engines such as Unity or Unreal Engine is that they allow the development and building of a video game for multiple target devices (PC, mobile, consoles, etc.). By targeting multiple platforms, developers can reach a broader user base. New differences between the builds for different devices in both engines may emerge, thus making the analysis of the same game io different platforms a goal for  
our future work.


\section*{Declaration of Competing Interest}
The authors do not have any conflict of interest.

\section*{Acknowledgements}
This work was supported in part by the following projects:

\begin{itemize}
    \item Project VARIATIVA (Ministry of Economy and Competitiveness
(MINECO) through the Spanish National R+D+i Plan and ERDF funds under Grant PID2021-128695OB-I00)
    \item  Research Group S05\_20D. Gobierno de Arag\'on (Spain).
    \item OASSIS (PID2021-122554OBC31/AEI/10.13039/ 501100011033/FEDER, UE)
    \item EMMA (Project SBPLY/21/180501/000115, funded by CECD (JCCM) and FEDER funds)
    \item SEEAT (PDC2022-133249-C31 funded by MCIN/AEI/ 10.13039/501100011033 and European Union NextGenerationEU/
    \item PLAGEMIS (TED2021-129245B-C22 funded by MCIN/AEI/ 10.13039/501100011033 and European Union NextGenerationEU/PRTR)
\end{itemize}


 \bibliographystyle{elsarticle-num} 
 \bibliography{references}





\end{document}